# More Than Peer Production: Fanfiction Communities as Sites of Distributed Mentoring

Sarah Evans, Katie Davis, Abigail Evans, Julie Ann Campbell, David P. Randall,
Kodlee Yin, Cecilia Aragon
University of Washington, Seattle, USA
{sarahe,kdavis78,abievans,juliemu,dpr47}@uw.edu, kodlee@fru1t.me, aragon@uw.edu

**ABSTRACT**
From Harry Potter to American Horror Story, fanfiction is extremely popular among young people. Sites such as Fanfiction.net host millions of stories, with thousands more posted each day. Enthusiasts are sharing their writing and reading stories written by others. Exactly how does a generation known more for videogame expertise than long-form writing become so engaged in reading and writing in these communities? Via a nine-month ethnographic investigation of fanfiction communities that included participant observation, interviews, a thematic analysis of 4,500 reader reviews and an in-depth case study of a discussion group, we found that members of fanfiction communities spontaneously mentor each other in open forums, and that this mentoring builds upon previous interactions in a way that is distinct from traditional forms of mentoring and made possible by the affordances of networked publics. This work extends and develops the theory of *distributed mentoring*. Our findings illustrate how distributed mentoring supports fanfiction authors as they work to develop their writing skills. We believe distributed mentoring holds potential for supporting learning in a variety of formal and informal learning environments.

**ACM Classification Keywords**
H.5.3 Group and organization interfaces: Web-based interaction.

**INTRODUCTION**
The way we ask for and receive help has changed with the emergence of networked technologies. Instead of opening up an encyclopedia, we conduct a Google search or consult Yahoo! Answers. Instead of searching the yellow pages, we look at Yelp reviews. Instead of asking a neighbor for a referral, we post on Facebook. Our sources of guidance are no longer singular, static voices of authority, but rather multiple, distributed voices, each with a distinct perspective [23,47]. Online, these sources come in the form of social Q&A sites [23], crowdsourced feedback on social network platforms [26], online critiquing communities [43,52], and formal and informal mentoring platforms [1,48,51].

This paper focuses on informal mentoring processes that take place in online fanfiction communities, which bring people together around a shared passion for a particular fandom (e.g., Harry Potter, Doctor Who) and in which people share and critique each other's amateur fanfiction writing. Our work is distinct from investigations of social Q&A sites and crowdsourced feedback [e.g., 23,26], which are not explicitly focused on personal development or sustained relationships around a shared interest. Though online critiquing sites do include these attributes [43,52], they tend to be more focused on professional development than online fanfiction communities, which are first and foremost about a shared personal passion [35,36]. Although there is quite a lot of research on Internet-enabled mentoring, most of this work looks at formal mentoring programs rather than spontaneous mentoring in online affinity spaces [16,48,51]. Studies that address the topic of informal mentoring in online affinity spaces [e.g., 1,9,10, 34,36,53] do not focus on the theoretical underpinnings of online mentoring and its relationship to traditional models of mentoring. In the current study, we draw explicitly on this prior theoretical work [e.g., 8,14,25,27,28] in our investigation of *distributed mentoring* [11] in online fanfiction communities.

We build on and extend previous work by Campbell et al. [11] that examined mentoring relationships in the context of online fanfiction communities. This research identified mentoring processes that were uniquely supported by the affordances of networked technologies. This new form of network-enabled mentoring is called *distributed mentoring*. Grounded in Hutchins' theory of distributed cognition [30,31] and directly tied to the particular qualities of networked publics, distributed mentoring is defined by seven key attributes that distinguish it from traditional, offline forms of mentoring: *aggregation, accretion, acceleration, abundance, availability, asynchronicity,* and *affect.* Through these attributes, authors gain valuable feedback on their fanfiction that they use to improve the quality of their writing.

Previous work resulted in a detailed description of the characteristics of distributed mentoring with examples and evidence gleaned from author interviews and participant observations [11]. The current study expands the theory further by revealing its operation across a larger dataset, as well as how it plays out on a moment-by-moment basis in fanfiction communities. In the process, we demonstrate the substantive differences between distributed mentoring and traditional forms of mentoring, as well as online forms of peer production. We also explore the central role that networked technologies play in giving rise to distributed mentoring. Through a thematic analysis of 4,500 fanfiction reader reviews posted in the online repository Fanfiction.net, we document the prevalence of different types of feedback as well as other types of interactions occurring in these reviews. This analysis provides quantitative evidence of distributed mentoring occurring across a range of fandoms and fanfiction stories. To complement this broad investigation of distributed mentoring, we conducted an in-depth case study of a single online writing support group dedicated to My Little Pony fanfiction. We provide a detailed account of one particular group discussion to explore "up close" how fanfiction participants interact with each other, building on each other's ideas in an asynchronous, cumulative fashion.

The contribution of this paper is an extension of the previous study, focusing explicitly on how the distributed nature of knowledge sharing transforms mentoring processes in distinct ways, constituting further empirical evidence for distributed mentoring and a detailed examination of its structure within the context of a prototypical example of participatory culture—fanfiction. Specifically, the current investigation provides two distinct and complementary views of how the attributes of distributed mentoring manifest and work together to support fanfiction authors in developing their writing skills. We discuss the application of distributed mentoring to other contexts and identify avenues for future work to investigate the relationship between distributed mentoring and learning in additional networked communities.

**PREVIOUS WORK AND BACKGROUND**

**Networked Participation**
Because of their ability to connect people across space and time, networked technologies introduce new opportunities for participation and learning [7,22,33,34,36,37]. Jenkins introduced the term *participatory culture* to describe "a culture with relatively low barriers to artistic expression and civic engagement, strong support for creating and sharing creations, and some type of informal mentorship whereby experienced participants pass along knowledge to novices" [37]. Participatory cultures center on the ability to create and share one's creations with others, something that networked technologies are well positioned to support. Sites of participatory culture include social media platforms like Facebook, Tumblr, and Instagram; gaming worlds like League of Legends and Minecraft; and knowledge-building sites like Wikipedia. A key attribute of participatory culture is the social connection that participants feel as they engage with each other in activities that matter to them. In our work, we focus on mentoring relationships, a specific type of social connection supported by participatory culture. Jenkins describes the role of informal mentoring in participatory culture as a means for experienced members to pass on knowledge to less experienced members. We build on this description by delineating the specific characteristics of informal mentoring found in a particular hotbed of participatory culture: online fanfiction communities.

Where there is participatory culture, there are affinity spaces. Gee [20] uses the term *affinity spaces* to describe communities that are defined by interest-driven participation. People come together around a shared passion, and their subsequent participation in the community is driven by that passion. Like Jenkins, Gee describes the informal mentoring that spontaneously arises in affinity spaces. As a core means for deepening novice members' participation in a social practice, informal mentoring represents a key mechanism for learning in affinity spaces [21,38]. Newer members gain knowledge and expertise through guided participation from more experienced members. Gee underscores the distributed, dispersed, and tacit nature of knowledge that is created and shared in affinity spaces. We extend this work by focusing on how distributed mentoring goes beyond peer production to enable a rich and diverse network of feedback that fanfiction authors use to inform and improve their writing. In addition, whereas affinity spaces can be either virtual or physical, our model of distributed mentoring focuses explicitly on mentoring that occurs in online environments and takes advantage of the distinct affordances of networked technologies.

**Participation in Fanfiction Communities**
As a particular type of affinity space, fan communities represent fertile ground for participatory culture to flourish. Indeed, Jenkins' seminal explorations of fan culture and communities provided the foundation for his articulation of the concept and qualities of participatory culture [35]. He described how fanfiction authors actively engage with source material and each other, extending original works in new and creative ways. Building on Jenkins' work, researchers have documented the literacies that youth develop through their participation in writing and critiquing fanfiction [2,3,4,12].

Fanfiction originally developed in the 1960s within fandoms for television shows like Star Trek [35], although some fanfiction authors point to the works of Shakespeare and the Aeneid as examples of early fanfiction. During the early decades of fanfiction, the primary mode for authors to receive feedback was through printed zines. Zines gave readers the opportunity to respond to fanfiction stories with "Letters of Comment," where feedback was provided on topics such as plot, writing style, and canon adherence [35].

The advent of networked technologies has had a notable impact on the means of producing, distributing, and engaging with fanfiction. The online reader review represents the Internet age descendant of the zine's Letters of Comment, lowering the barriers to providing feedback due to the absence of the physical and temporal constraints associated with print. The online format also supports instant, highly interactive communications among reviewers and fanfiction authors. As a result, there is now a considerably higher volume of information generated by and shared among larger numbers of participants compared to in pre-networked times. This new information ecosystem supports a rich network of support, commentary, communication, and mentoring: *distributed mentoring*.

Online fanfiction communities are a type of affinity group that is well-suited to mentoring activities. Participants on such sites are by definition amateur writers who come to experiment and receive feedback from interested readers. Unlike a published book, which has gone through an extended editing and revision process, a fanfiction site is a place to workshop new material, and reviewers are conscious of being a part of the process. In Campbell et al.'s interviews [11], all authors described a need to participate in the review process for others, giving back to the community.

**Forms of Mentoring**
Although mentoring has likely occurred throughout human history, related research and professional literature on the topic did not appear until the 1970's [32]. Since that time, authors have wrestled to define and theorize mentorship. Bozeman and Feeney [8] critiqued 20 years of research to create a narrow definition of mentoring that requires a dyad focused on career or personal development. The authors felt that activities falling outside their scope should be given a different term such as "coaching" or "socialization." Bozeman and Feeney's narrow definition is limiting and fails to include divergent mentoring situations where participants receive instrumental and psychosocial support. For example, Lave and Wenger [41] describe several apprenticeship situations where peer mentorship in groups played a critical role in skill development and knowledge accumulation.

Subsequent work by other authors reveals the narrowness of previous mentorship definitions and continues to explore the boundaries of what creates a mentoring situation. For example, Huizing [27] pushed on this definition to allow for group mentoring experiences in four subtypes: peer group, one-to-many, many-to-one, and many-to-many. Kroll [40] called for a formal definition of group mentorship that includes specific and shared purposes to challenge and support others. Additional researchers have focused on factors such as identity [28] and personal connection [50] to draw boundaries around mentorship.

To account for the variety of scenarios labeled mentorship, Dawson [14] offers a framework of sixteen dimensions along which researchers and practitioners can position their findings. Developed from a review of the literature and feedback from mentoring professionals, Dawson more accurately accounts for the range of situations from which people report mentoring experiences. The dimensions identified are objectives, roles, cardinality, tie strength, relative seniority, time, selection, matching, activities, resources and tools, role of technology, training, rewards, policy, monitoring, and termination. The structured yet flexible nature of these dimensions allows for more descriptive definitions of the mentoring that occurs in many real-life situations, both on and offline. While these dimensions vastly improve the ability of researchers to compare and contrast mentorship activities, the framework is not without limitations. In particular, the affordances of networked technologies [16] cannot be fully accounted for without additional theorizing.

**Distributed Mentoring**
Distributed mentoring is grounded in Hutchins' concept of *distributed cognition* [30,31]. Useful for understanding the nature of group activity in a variety of contexts, including affinity spaces, distributed cognition represents a framework for understanding how complex tasks are completed through distributed processes of coordinated activity. Instead of the individual, the group and its environment represent the primary unit of analysis, including the tools available in the environment and the uses to which they are put. Hutchins showed how individuals become part of a cognitive system, and that the cognitive tasks made possible through the coordination of people, tools, and actions are greater than what any one individual or tool could accomplish on its own. Distributed mentoring builds upon this understanding of emergent systems and extends it beyond the cognitive aspects.

Though Hutchins' initial investigations were conducted in offline environments, distributed cognition has broad applicability to online spaces. The affordances of networked technologies—including asynchronous communication, easily searchable content, and the ability to reach a wide audience—contribute to the distributed nature of knowledge in online environments and facilitate its coordination. Campbell et al. [11] observed how these affordances play a role in shaping mentoring processes in online fanfiction communities. This analysis led to the articulation of distributed mentoring as a network-enabled form of mentoring in which fanfiction authors receive feedback and encouragement from a wide variety of people using diverse tools and platforms. The resulting knowledge generated through the aggregation of this feedback is greater and qualitatively different from the knowledge residing in any one individual or artifact.

As in systems of distributed cognition, distributed mentoring is a cognitive ecosystem where information, in this case information useful to the development of writing skills, is embodied in artifacts of the system, including reviews, private messages, and group discussions. While many other affinity spaces contain similar communication features, in fanfiction communities these artifacts operate together systematically to support and teach writers. Individual readers and writers contribute their portion of knowledge to the system, be it guidance on crafting characters or grammar tips, and in return receive all the wisdom of the system shaping their writing practice. Because participants in online fanfiction communities are there because of their shared love of fanfiction, the feedback and guidance that writers receive is intended not just to improve the work but also the person producing the work. It is this sense of community, based on shared interests, that distinguishes fanfiction communities from other sites of feedback and critique and turns them into rich sites of distributed mentoring.

Distributed mentoring represents a notable departure from traditional models of mentoring, which typically emphasize formal selection and matching of mentor-mentee relationships, the importance of relative seniority, clearly delimited time commitments, and even mentorship training [14,28,29,39,42]. These models expect mentoring activities to be clearly labeled as such and often assume that the mentorship focuses exclusively on career issues [8]. In contrast, distributed mentoring is defined by its fluid, non-hierarchical relationships that form and dissolve in an impromptu manner, each one contributing just one portion of the overarching mentorship that authors experience when they pool the many sources of feedback they receive. In this way, distributed mentoring is more closely aligned with recent, expanded definitions of mentoring, which recognize that mentoring relationships can take place between participants of all ages, with different levels of engagement and formality, and in varying contexts [14]. Though the configurations and contexts may look different, the primary function of distributed mentoring remains consistent with traditional models of mentoring defined by previous research: to provide instrumental support (such as providing knowledge or social capital) and psychosocial support [8,15,25,29].

Distributed mentoring is defined by seven key attributes that illustrate its distinctness from traditional, non-networked forms of mentoring as well as its unique benefits [11]. Through the *aggregation* of feedback from diverse members of the fanfiction community, authors understand what they have done well and what is lacking in their fanfiction stories. Reviewers of fanfiction stories themselves interact with each other in a persistent, cumulative manner, enabling an *accretion* of knowledge to facilitate the author's learning process. The *acceleration* of knowledge and learning is facilitated by the constant connectivity of networked technologies, which supports ongoing, active discussions among reviewers. These discussions produce rich feedback for authors as reviewers push back on each other's points or expand on each other's comments. The sheer *abundance* of review responses represents another attribute of distributed mentoring, since a large volume of feedback, however shallow each comment might be (*e.g., "I love it!"*), provides overall direction to the writer. The persistent and public nature of online text-based communication makes possible the long-term *availability* of reviews, which facilitates sustained exchanges and relationships among reviewers and authors. The *asynchronicity* of networked communication allows for the transcendence of time and space, enabling reviewers across the globe to view and reply to other reviews easily and continuously. Lastly, fanfiction writers underscored the positive *affect* or emotion they experienced through reviewers' encouraging comments. These seven attributes of distributed mentoring work together—often in an overlapping manner—to deepen and enrich the mentorship that authors receive through their participation in online fanfiction communities. In addition, the seven attributes distinguish distributed mentoring from Gee's [20] affinity spaces framework, whose lack of explicit emphasis on the distinct affordances of networked technologies limits its ability to account for the full range of mentoring activities we observed in our investigation.

Campbell et al. [11] introduced the concept of distributed mentoring, but focused on definition and theory building. In this paper, we further explore systematically the mechanisms of distributed mentoring, focusing on a detailed thematic analysis of the primary public artifacts of the distributed mentoring process in the fanfiction community—reviews—and an in-depth case study deconstructing fanfiction discussion on a single-fandom forum. This empirical approach enables us both to document systematic evidence of the existence of distributed mentoring and elucidate more details of its mechanisms.

## METHODS

We conducted a nine-month ethnographic investigation of online communities centered on three fandoms: Harry Potter, Doctor Who, and My Little Pony: Friendship Is Magic. We conducted in-depth interviews with 28 young fanfiction authors (ages 13-30, mean age 22.8 years), focusing on the different forms that mentoring takes in online fanfiction communities. We explored these themes in situ through participant observations in a variety of online communities dedicated to fanfiction related to our three focal fandoms. These observations allowed us to document specific processes of interpersonal communication among members of the fan communities.

Following this broad investigation, we narrowed our focus in order to investigate systematically the prevalence and mechanisms of distributed mentoring. We conducted a thematic analysis of 4,500 fanfiction reader reviews and an in-depth case study of an online writing support group dedicated to My Little Pony fanfiction. Our analysis of reader reviews provides quantitative evidence of distributed mentoring on a large scale, while the case study provides a detailed account of distributed mentoring in action.

The following research questions guided our investigation:

1. *To what extent do the reader reviews of fanfiction stories include evidence of distributed mentoring?*
2. *What does distributed mentoring look like in a single online fanfiction community?*

Factors affecting our selection of the three fandoms included the research team's personal familiarity with the fandoms and our desire to ensure diversity with respect to genre, medium, and length of time in existence. The inclusion of three fandoms in our thematic analysis allowed us to investigate similarities and differences in the nature of distributed mentoring across disparate fandoms.

**Study Sites**
Fanfiction.net and FIMFiction.net were the primary sites for our research. Both websites are fanfiction repositories, which allow authors to post stories and receive reviews of their work. We conducted our thematic analysis of reader reviews in Fanfiction.net because it is the largest and most popular fanfiction repository on the Internet. The site has over seven million registered users and hosts more than five million fanfictions across thousands of fandoms. Users post stories they have authored, and other registered users can choose to write reviews for these stories. Frequently, authors request specific feedback in their author's notes at the start of a story. To understand the scale of activity on Fanfiction.net, we wrote automated scripts to scrape site data. For the three fandoms explored in our study, 190,364 authors had posted 511,726 stories as of May 2016. The number of reviews that individual stories received ranged from 0 to 31,863, with an average of 26 reviews per story.

Concurrently with our thematic analysis, we conducted a case study of an online writing support group in FIMFiction.net, a single-fandom fanfiction repository dedicated to the television show My Little Pony: Friendship is Magic. Considerably smaller than Fanfiction.net, FIMFiction.net had approximately 210,000 registered users and 98,185 published stories as of May 2016 [19]. In addition to individual user pages, FIMFiction.net features more than 7,000 user groups. These groups include a common fanfiction repository and discussion forum, and they cover a range of topics related to specific aspects of the fandom (*e.g.,* favorite characters) and the craft of writing (*e.g.,* how to create a compelling antagonist).

**Participant Observation and Interviews**
Over the course of our nine-month investigation, each member of the research team participated in at least one of the communities being studied. This participation included writing reviews, general commenting behavior, up/down voting and rating stories, and each writing one or more chapters of fanfiction. Several of the researchers also published this writing in an appropriate venue for feedback, providing us with first-hand experience of not only the writing and publishing process across different sites, but also the process of receiving critical feedback and mentoring from members of the fanfiction communities. These mentoring experiences were critical in helping us to gain a deeper understanding of the communities that were being studied, and in formulating the question protocols for our interviews with authors.

During our participant observation, each member of the research team identified authors across the communities who were active members (regularly posting stories and reviews) and between the ages of 13-30 (adolescents and emerging adults). We then reached out to these individuals through email or private message and invited them to participate in an interview about their experiences writing fanfiction. These interviews were conducted online through an asynchronous series of either email or private message exchanges between the participant and a single researcher. In order to break up the question protocol—and to allow researchers to adapt based upon responses—the full set of interview questions required three message exchanges to complete. We contacted 32 individuals and received responses from 28. Members of the My Little Pony community represented the bulk of respondents (12), followed by Doctor Who (9) and then Harry Potter (7).

**Thematic Analysis of Fanfiction Reviews**
Many of our interview participants stressed how important fanfiction reviews were to them, both as a motivating force and as a learning experience. In our participant observations, we experienced firsthand the emotional impact of receiving comments on our own creative writing and the valuable advice and support provided in fanfiction reviews. These insights inspired us to explore reviews on a larger, more systematic scale, conducting a deeper, thematic investigation of 4,500 reader reviews to develop an understanding of the prevalence of different types of feedback as well as other types of interactions occurring in reviews.

As the first step in our thematic analysis [6], we developed an initial "start list" of codes [45] after reading through the complete set of 133 reviews for a single Harry Potter fanfiction story written by one of the authors we interviewed. This approach allowed us to engage the author in direct discussion about the types of feedback that helped her writing, thus ensuring that our emerging codes were grounded in participants' lived experiences. We expanded and refined our start list of codes by reading through 777 reviews from six additional fanfiction stories from Doctor Who and Harry Potter. Though our coding scheme bears resemblance to coding schemes used in other studies [e.g., 2,13,18], we tailored it specifically to the themes that emerged directly from our specific dataset [45].

After developing our initial set of codes, we completed four rounds of trial coding. During each round, five members of our research team coded a set of reviews independently. We recorded areas of agreement and disagreement for each review, discussed disagreements in weekly team meetings, and arrived at a group consensus for each review. We also refined our coding scheme so that it reflected the full range

|   | **Code** | **Description** | **κ** | **Occurrence** | **% of Reviews** |
|---|---|---|---|---|---|
| 1 | Shallow positive | Positive reviews that do not provide specific feedback about the text. | 0.89 | 1580 reviews | 35.1% |
| 2 | Targeted positive | Reviews positively reflecting on specific aspects of the text. | 0.79 | 1351 reviews | 30.0% |
| 3 | Targeted corrective or constructive | Critical or neutral feedback on specific aspects of the text, *e.g.,* grammar and plot suggestions. | 0.75 | 747 reviews | 16.6% |
| 4 | Targeted positive and corrective/constructive | Both sets of feedback must call out specific aspects of the text described in 2 & 3. | 0.72 | 243 reviews | 5.4% |
| 5 | Non-constructive negative | Troll posts or flames where the reviewer is intentionally antagonizing the author. | 0.79 | 45 reviews | 1.0% |
| 6 | Discussion about the story | Reviewers or authors replying to or referencing each other when discussing the story or starting a discussion by asking questions about the story. | 0.94 | 389 reviews | 8.6% |
| 7 | Discussion not about the story | Reviewers or authors discussing topics unrelated to the story, *e.g.,* daily life. | 0.71 | 86 reviews | 1.9% |
| 8 | Fandom remarks | Reviewers drawing on canon or fanon (fan canon) to position themselves with regard to their fan knowledge. | 0.81 | 466 reviews | 10.4% |
| 9 | One-sided connection | Comments suggesting an ongoing relationship on the reader's side, *e.g.,* following the author's collective work. | 0.34 | 175 reviews | 3.9% |
| 10 | Two-sided connection | Comments suggesting an ongoing relationship between the reader and the author. | 0.25 | 55 reviews | 1.2% |
| 11 | Review fishing | Reviewers asking for reviews on their own fanfictions. | N/A* | 13 reviews | 0.3% |
| 12 | Update encouragement | Encouraging the author to write more. | 0.88 | 1240 reviews | 27.6% |
| 13 | Miscellaneous | Undecipherable text or otherwise uncategorizable. | N/A* | 74 reviews | 1.6% |

*Code did not occur frequently enough to measure.

Table 1. Code name and description, inter-rater reliability statistics (Fleiss' Kappa), code occurrence, and percentage of reviews (total reviews = 4,500) that included each code.

of comments we observed in the reviews. This process of collaborative coding [49] ensured that we applied the codes consistently and accurately to our data.

Next, we randomly sampled 4,500 reviews from several categories on Fanfiction.net. We organized the fanfiction stories into high-, medium-, and low-popularity categories based on the number of reviews they had received. The high-popularity category contained fanfiction stories that were in the top 0.5 percentile of stories in terms of total number of reviews. The medium-popularity category included stories in the 0.5 to 10th percentile range, and the low-popularity category included stories in the 10th to 50th percentile range (stories below the 50th percentile had few to no reviews). We sampled from these three ranges to ensure representation of reviews from fanfiction stories with different levels of popularity, in case review types varied with story popularity. We selected fanfiction stories based on these percentiles because the distribution of stories by number of reviews followed a power law, with stories below the 50th percentile having fewer than five reviews on average, while stories in the top 0.1 percentile had more than 1000 reviews on average. The final coding scheme encompassed 13 codes (Table 1). Each review could have more than one code applied to it, but codes 1, 2, 3, and 4 were mutually exclusive. Because we had five team members involved in the coding process, we used Fleiss' Kappa to measure inter-rater reliability [17]. For each review, the team members recorded whether or not it contained each of the 13 individual codes, using a Boolean notation system. Inter-rater reliability was consequently calculated per code, so a single review could pertain to multiple reliability ratings. Table 1 shows the codes and their descriptions along with the measure of inter-rater reliability for each code. For all codes except 9, 10, 11, and 13 the Fleiss' Kappa (κ) values were between 0.71 and 0.94, representing excellent agreement. Codes that did not produce excellent agreement occurred very infrequently, which is a common cause for low inter-rater agreement. To ensure that these codes were applied consistently and accurately throughout the entire data set, a second coder verified reviews involving these codes.

To begin our analysis of the coded set of reviews, we totaled the codes for each category (high-, medium-, and low-popularity) and fandom (Harry Potter, Doctor Who, My Little Pony) to determine which codes were most frequently applied. We converted the total into a proportion based on the number of reviews per category in order to facilitate comparisons among the high-, medium-, and low-popularity fanfiction categories and among the three fandoms. We looked for patterns in the co-occurrence of codes, and we conducted chi-square tests to check for significant systematic variations in the proportions of reviews among popularity categories and fandoms. We have not broken out the code occurrences by fandom because our chi-square tests revealed no systematic differences across the three fandoms. For popularity categories, we report only those chi-square tests that were statistically significant (Table 2). Finally, we examined what different rates of occurrence among the codes meant in terms of the mentoring being provided in reviews and the type of community these reviews reflected.

**Case Study Analysis**
Concurrent with our thematic analysis, we selected for the focus of our case study the largest and most active writing group on FIMFiction.net. This group is a general writing community where authors ask for advice about specific problems they are facing in their writing. During the course of our author interviews and participant observations of five writing groups, the high degree of participation and lively conversation among users in this particular FIMFiction.net group prompted us to investigate it in greater depth.

Over a five-month period, a member of our research team averaged two hours per week observing the FIMFiction.net writing group, totaling approximately 40 hours of participant observation. The researcher's documentation of her observations included field notes and screen captures. Additionally, she communicated informally via private message with members of the group to ask them questions about their participation. Due to this researcher's long history of participation in online fandom and fanfiction communities, extensive knowledge of the My Little Pony fandom, and prior experience posting her own fanfiction stories on FIMFiction.net, she was well positioned to engage in the writing group and interpret the interactions she observed [2,5]. The researcher shared her observations and field notes with the rest of the team members in weekly meetings, which provided a forum for identifying notable posts, comments, and interactions among community members [24,44,49]. During the course of these weekly meetings, the research team identified the posts described in this paper as a particularly generative discussion that illustrates the seven attributes of distributed mentoring.

Quotations included are unmodified, unless obfuscation was necessary for anonymity, and as a result may contain typographical errors that were present in their original form.

**FINDINGS**
Our thematic analysis of the reviews revealed distinct patterns of communication and feedback that collectively illustrate the nature of mentoring through reader reviews on Fanfiction.net. The coded reviews were overwhelmingly positive, and over 50% contained substantive feedback (instrumental support). We also documented evidence of ongoing relationships among authors and readers (psychosocial support). Our analysis of a single discussion in one FIMFiction.net group provided an in-depth look at how community members interact with each other and build on each other's comments in a manner quite distinct from traditional one-to-one forms of mentoring. In the findings reported below, we list the related attributes of distributed mentoring alongside the relevant results.

**Mentoring through Story Reviews**
Fanfiction reviews present a fertile resource for authors, where they can learn from the tens, hundreds, or even thousands of reviews left on their work. Due to the site's large user base, fanfiction stories posted on Fanfiction.net may receive a large number of reader reviews. Because interview participants pointed to these reviews as a key source of feedback, we decided to investigate the mentoring processes in these reader reviews in greater depth. Table 1 presents a summary of the code occurrences and the percentage of reviews in our sample that included that code.

Across the three fandoms in our study, shallow positive comments (*e.g., "Incredible"*) represented the most commonly occurring code in our sample of reviews, with 35.1% of the reviews containing this type of comment (abundance, affect). Targeted, positive feedback was included in 30.0% of the reviews, while update encouragements were part of 27.6% of reviews. Notably, the combined category of targeted feedback, which comprises purely positive comments, constructive criticism, and a mixture of positive comments and critiques, occurred in more than 50% of the reviews. This finding indicates that a substantial portion of reviews went beyond a simple shallow response to offer authors substantive, instrumental feedback on their stories (aggregation). Targeted, positive feedback called out specific positive aspects about the text, *e.g., "I like the ambiguity of your ending, it leaves me feeling hopeful."* Another 22% of the reviews contained targeted feedback that was either constructive (negative) or contained both positive and negative elements, *e.g., "I think this could have been edited massively to reduce passive voice…Still, this gets my upvote."* Our interview participants described the collective value they derive from this variety of feedback:

*The brief positive reviews probably make up the majority, and I don't tend to dwell on them very much, though obviously they're very nice reviews to receive. The more specific ones make a little more of an impact, they usually refer to something I was particularly pleased with or something I felt was harder to convey… (Author 16, Harry Potter)*

Participants reflected explicitly on the learning value they derive from reviews: *"Yes, writing fanfiction and getting*

*instant feedback over the past couple of years has improved my writing significantly"* (Author 19, My Little Pony) (availability, asynchronicity). Even if the feedback sometimes was unpleasant to receive, participants said they still appreciated and learned from it:

*The moment you realise the reviewer has a point is when your motivation takes a hit. But I believe if writing is something you're passionate about, you can come back from it [the criticism] and be better for it. My usual response for such reviews is to message the reviewer and ask them to elaborate and be specific in what they didn't like. (Author 14, Harry Potter)*

The least frequent code was review fishing, with only 13 incidents of review fishing recorded. The low incidence of review fishing fits with the spirit of reciprocity in the fanfiction community visible in comments made by the authors we interviewed. The authors indicated that they knew how much reviews were valued and stated that they wanted to return the favor by leaving reviews for others (aggregation, accretion, acceleration); therefore, they may not need to be reminded to review.

The number of positive messages in reviews far outweighed the number of negative messages. Shallow positive and targeted positive comments occurred in 70.5% of reviews (abundance, affect), while only 1.0% of reviews contained non-constructive negative comments like flames, *e.g., "I never thought that human spawn could create such a horrible piece of crap."* This finding illustrates the positive atmosphere of fanfiction communities (affect). We even observed instances where reviewers replied to flame posts in an effort to defend the author of the story being criticized (acceleration, accretion). One controversial fanfiction story that we analyzed contained several flame reviews but even more reviews like this one defending the author:

*Do you realise what you've started? It's like a war between all of the fans who hate Reinnette or enjoy this story and those who have their heads stuck up their butts and have nothing better to do than be rude about this fic. Quite frankly screw them, and good for you, because I think you're going to go down in fanfic history for this! (Doctor Who)*

In 10.4% of reviews, readers added fandom remarks based on canon or fanon insights. In these types of comments, reviewers identified themselves as part of the larger fandom community by expressing great enthusiasm for aspects of the fandom or by demonstrating deeper knowledge of the canon or fanon material beyond what was contained in the story reviewed (accretion, acceleration). Many fandom remarks contained corrective feedback for authors who made mistakes with regard to the canon, *e.g., "I'm sure it's obnoxious, but I feel the need to point out that Slytherin's seeker in Harry's first year was Terence Higgs, not Marcus Flint. Flint was always a chaser."*

Because mentoring is inherently relational, we used our coding scheme to document interactions among readers and authors (accretion, acceleration), as well as evidence of ongoing relationships among authors and readers (availability, affect). Reviews containing back-and-forth discussion about the story were found in 8.6% of the story reviews (accretion, acceleration), whereas discussion unrelated to the story occurred much less frequently (1.9% of all reviews). Responses to author's notes were a common type of discussion included in reviews. One author left a note warning readers that they may be unsatisfied with the recent chapter update, writing: *"You may find this chapter frustrating. Never fear—the next chapter will be up on Tuesday, and you will have answers."* To this author's note a reviewer responded: *"You were right. I want to smack both of them. And I have got to say, Hermione better not lose that baby. I want to imagine Severus's face when he finds out lol :)"* This response showed agreement with the author's note and provided a plot suggestion to resolve the story in a satisfactory manner.

We coded reviews that suggested an ongoing, reciprocal relationship between the reader and author (affect), *e.g., "You're just getting out, too? Sweet. You know, I think I've got an idea down for a little side-shot like this. I'll get working on it ASAP and send it to you, see what you think."* The number of reviews showing this type of two-sided connection between an author and a reader (1.2% of all stories coded) was considerably lower than the number showing a one-sided connection (3.9% of all stories coded) in which the reader made a comment suggesting s/he followed the author's work with some regularity (affect, availability), *e.g., "The way you craft stories is breathtaking, it really is; you understand the characters better than anyone else I've seen... Here's to the stories written, and the stories yet to come."* The lower frequency of two-sided connections may be attributable to the fact that Fanfiction.net members tend to communicate via private message, something we learned from our author interviews. When a Fanfiction.net user clicks the "reply" icon next to a review, a private message screen opens that includes a quotation of the review. Several of the authors we interviewed stated that they reply to all of their reviews—or at least the reviews containing constructive criticism—via this direct reply system (asynchronicity, availability).

In order to understand more about the network structure of reviews, especially in relation to the more popular stories, we conducted chi-square tests over frequency of each code type across stories at the high-, medium-, and low-popularity levels. This analysis allowed us to determine whether the frequencies of each code reported above differed according to the story's popularity. Across all 13 codes, we found very few statistically significant differences among high, medium, and low levels of reviews. Two notable exceptions included the code measuring discussion about the story (code #6) and the code measuring a one-sided connection with the author (code #9), which each showed a statistically significant difference in frequency based on a chi-square test ($\chi2 < 0.05$).

Table 2 displays the occurrence for these two codes at the high-, medium-, and low-popularity levels. Both codes occurred most frequently at the high-popularity level, falling at the medium-popularity level and again at the low-popularity level. These results show that there is a greater degree of connectedness among reviewers and authors for more popular fanfiction stories (accretion, acceleration, affect). This finding was corroborated by our author interviews: *"I also have a lot of friends who are happy to preread or edit for me, point out reoccurring mistakes and ways to improve my writing and my story" (Author 23, MLP).*

|  | Code #6 Discussion about the story | Code #9 One-sided connection |
|---|---|---|
| $\chi^2$ | 7.11 | 8.12 |
| $p$ | 0.03 | 0.02 |
| Percent of high reviews | 9.60% | 4.84% |
| Percent of medium reviews | 8.27% | 3.60% |
| Percent of low reviews | 6.53% | 1.60% |

Table 2. Code comparison across stories that received a high, medium, and low number of reviews.

**Case Study of a FIMFiction.net Group**

Our case study allowed us to examine in greater detail a specific instance of distributed mentoring. The group on FIMFiction.net, where authors can ask for advice, provides learning opportunities both for writers asking questions as well as group members reading and responding to the questions. With dozens of replies posted each day, the conversations in this group move fast, and the vast majority of posts receive multiple replies (abundance, acceleration, availability). Forum posts on FIMFiction.net do not have threaded replies, so members reply to each other by indicating the name of the users to whom they are directing their comments using double angle brackets, >>. Because of this lack of threading, members are able to make a single post referencing multiple other respondents (accretion, acceleration, asynchronicity). The post described here is representative of many of the question-and-answer posts from this group. In this post, the original poster asked for advice on how to write the character Princess Luna, an antagonist from the first season of My Little Pony: Friendship is Magic. The post received 46 responses from 20 unique respondents. Eight of the respondents simply replied directly to the original poster without engaging other respondents in conversation, while the remaining 12 respondents replied to one or more other respondents. Two respondents did not reply to the original poster at all, instead replying only to other respondents.

The respondents who engaged others in conversation corrected inaccurate information, supported useful suggestions, and debated differing opinions (accretion, acceleration). For example, Respondent 5 erroneously stated that Luna spoke in Old English, a common misconception. Respondent 6 replied, stating: *"It's important to note that Luna doesn't speak in Old English. We wouldn't be able to understand anything she says otherwise."* Another respondent (Respondent 14) referenced this point when referring to Luna's English as *"Early Modern English,"* which is a more accurate characterization of her language.

Respondents also indicated agreement with each other, letting the original poster know that the piece of advice was valid (acceleration). Respondent 1 suggested: *"From the limited amount of stories that I have read, Luna is usually portrayed as a gamer or somewhat out of touch with modern culture."* To this suggestion, Respondent 6 agreed, replying: *"While I'm picky about the kind of technology that I would introduce into a story, Luna being behind the times is right on the money."* This response also provided a qualification for the original poster to keep in mind when using the characterization that Luna is behind the times technologically: to be mindful of the technology introduced in fanfiction (accretion).

A short debate broke out between Respondent 16 and Respondent 19 about when different forms of speech were appropriate for Luna based on her backstory as a character. In this debate the respondents both showed their deep fandom knowledge and ended the discussion in a civil fashion (acceleration, affect):

*Respondent 19: >> [Original Poster] honestly, Luna doesn't even talk in that voice at this point.*

*Respondent 16: >> [Respondent 19] She said it's going to take place before her banishment.*

*Respondent 19: >> [Respondent 16] We already saw that she talked in her normal voice before the banishment.*

*Respondent 16: >> [Respondent 19] Unless her alter-ego was already in the driver's seat there. She always spoke 'normally'.*

*Respondent 19: >> [Respondent 16] yes, so why change it?*

*Respondent 16: >> [Respondent 19] Just because her alter-ego, a potential outside possessor spoke normally, doesn't mean Luna did? Conversely, obviously Luna spoke like that at least some of the time, when she was being formal. It might have been following out of favor colloquially. But you would still need formal speech for a fic with Luna as a real Princess. You're bound to run into it either way.*

*Respondent 19: >> [Respondent 16] She talked normally before she was changed into her alter-ego did you see the video?*

*Respondent 16: >> [Respondent 19] Yup. Watched it all the way through, since it remains awesome. But as I said, just because Luna turned into her alter-ego after that physically, doesn't mean she wasn't already in control mentally.*

*Respondent 19: >> [Respondent 16] I guess, or maybe the writers were just lazy*

*Respondent 16: >> [Respondent 19] It's a pretty good explanation for most anything.*

*Respondent 19: >> [Respondent 16] yep*

The public nature of group posts like this one allows for dissention and agreement among the respondents, thereby providing the original poster with a rich set of (persistent) advice to inform her story (availability). If the original poster had asked several people the same question privately and received differing opinions, she would not have had these interactions among the respondents to learn from (accretion, acceleration).

To investigate how authors used the diverse advice they received in these forum posts, we conversed informally with forum posters during the course of our participant observations, asking how they chose among the pieces of advice when writing their fanfictions. Authors described different strategies for selecting the best advice, including trying all of the pieces of advice and seeing what worked best; selecting the most repeated pieces of advice; or combining aspects of different pieces of advice into a solution that suited them best. One author explained: *"I guess you could say I mix the advice. I try each solution one at a time to see what works and what doesn't."* Another author observed: *"I usually just look through the responses and pick whichever ones seem to work best for me. This does mean that I will often compile ideas from various people, and so far it's worked very nicely."* All strategies described by authors took advantage of the advice presented by multiple respondents (aggregation, abundance, availability).

## DISCUSSION

The current study furthers our understanding of distributed mentoring and its attributes, including how it is made possible by networked communication and the value that people derive from it. This research is informed by prior work on participatory culture [36,37] and affinity spaces [20], as well as the influence of networked publics on these modes of participation [7,35]. While this existing work discussed new possibilities for informal mentoring, missing from these accounts was a delineation of the specific characteristics of informal mentoring found in affinity spaces and how these characteristics are shaped in distinct ways by the affordances of networked technologies. Past literature has defined mentoring in narrow ways [8, 14, 27, 49] that do not account for the support and development evident in interest-driven online communities. To fill this gap, we presented evidence of the seven attributes of distributed mentoring documented through an empirical investigation of three fanfiction communities that builds upon and expands previous work [11]. Grounded in Hutchins' theory of distributed cognition [30,31], distributed mentoring describes the processes by which authors receive feedback on their writing from a wide variety of sources distributed across time and space, and addresses the emergent properties that enable individuals to become part of a mentoring system, for example through *aggregation* of reviews and the *acceleration* provided through conflict and discussion among reviewers. The value that authors receive from distributed mentoring in online fanfiction communities is consistent with prior work documenting the learning that occurs in these communities [2,3,4,12].

The seven attributes of distributed mentoring and their interconnections were visible in both the thematic analysis of reader reviews and the discussion group case study. Table 3 summarizes the attributes and provides examples of how they manifested in our data. With respect to our thematic analysis, the prevalence of shallow positive comments (*e.g., "Awesome story!"*) shows how the *abundance* of feedback from readers can collectively provide both support (*affect*) and direction to authors. When viewed in the *aggregate* alongside the more substantive reviews—which represented more than half of the reviews in our sample—this feedback from diverse readers helps authors to understand where their stories are succeeding and where there is room for improvement. Interestingly, the non-positive targeted feedback frequently generated a significant number of new reviews in response, often containing contradictory positions. This pattern demonstrated *acceleration* as constructive feedback led to rich discussions, a phenomenon we noticed also in our case study. Interview data supported the finding that the *aggregation* of shallow and substantive feedback could add up to meaningful mentoring experiences, and contribute to an *accretion* of knowledge [43].

Further reinforcing the positive *affect* of online fanfiction communities, only 1% of our sample comprised troll reviews. This finding is similar to a recent investigation of comments about projects shared in the Scratch community [18], and stands in sharp contrast to conventional wisdom that anonymous comments tend to be overwhelmingly negative [46]. Instead, we found that reviewers and authors are linked in a networked system where giving and receiving advice generates positive affect. Reviewers (mentors) as a group seek to create an atmosphere of psychosocial support, which means attempting to balance negative comments with positive ones. Online connectivity facilitates active and long-term discussions of this nature.

In addition to the content of reviews, our thematic analysis gave us insight into the interactions among readers and authors and how the attributes of distributed mentoring support these interactions. Examples of interactions included responses to author's notes, reviews containing back-and-forth discussion about the story, and comments suggesting an ongoing connection between reviewer and author. The asynchronicity of communication on Fanfiction.net made it possible for community members to view and respond to each other's comments easily and continuously, while the constant connectivity of networked communication *accelerated* the process of giving and receiving feedback [16]. The persistent, public nature of online discussion made knowledge widely *available* and allowed it to *accrete* over time [16]. Taken together, these findings reveal how the attributes of distributed mentoring work together to provide a distinct mentoring experience that is fundamentally tied to the affordances of networked technologies [7,11]. While networked technologies give it a distinct flavor, distributed mentoring still shares the foundational qualities found in existing mentoring models: sustained instrumental and psychosocial support that is focused on skill development [15,25].

| Attribute | Description | Example from our analysis |
|---|---|---|
| *Aggregation* | In the aggregate, small pieces of feedback from multiple, independent community members help authors to identify strengths and weaknesses in a whole greater than the sum of its parts. | Over 50% of the analyzed reviews went beyond a simple shallow response to offer authors substantive, instrumental feedback on their stories. |
| *Accretion* | Reviewers interact with each other, drawing upon and building on earlier reviews. | Reviews containing back-and-forth discussion about the story were found in 8.6% of the analyzed reviews. |
| *Acceleration* | Conflict and discussion among reviewers leads to a network of feedback embedded with rich knowledge about the fandom and writing. | Our case study of FIMFiction.net showed how participants interacted with each other by correcting inaccurate information, supporting useful suggestions, and debating differing opinions. |
| *Abundance* | The large number of reviews can increase the weight of even the shallowest of feedback. | Both the thematic analysis and the case study showed the large amount of feedback that participants exchanged on Fanfiction.net and FIMFiction.net. |
| *Availability* | Online text-based communication between authors and reviewers remain available long into the future, allowing participants and observers to continue to learn from these exchanges even after they become inactive. | Our case study showed how the public nature of the forum posts on FIMFiction.net provided the original poster with a rich set of (persistent) advice to inform her story. |
| *Asynchronicity* | The asynchronous nature of online text-based communication allows diverse authors and reviewers to engage in discussion even when synchronous collaboration would be impossible. | Our case study of FIMFiction.net showed how participants responded to each other over time on forum posts. |
| *Affect* | Positive comments and interactions provide authors with valuable emotional support and encouragement. | In our thematic analysis, 70.5% of the review feedback was in the form of positive comments. |

**Table 3. The seven attributes of distributed mentoring with select examples drawn from our analysis.**

The fact that the reviews written for more highly reviewed stories displayed a greater degree of connectedness among reviewers and authors suggests that those authors who are more established in the community are better positioned to benefit from distributed mentoring due to the fact that they are more likely to receive feedback from reviewers who know and follow their work. This finding is consistent with Xu and Bailey [52], who found that users with higher reputations were more active in making reciprocal critiques in an online community focused on digital photography.

The evidence from the reader reviews provided a systematic view of distributed mentoring at scale, whereas the discussion group case study probed more deeply into one specific instance of distributed mentoring. Through this in-depth analysis, we were able to document specific communication patterns among participants. As with the reader reviews, comments left in the FIMFiction.net discussion group were *asynchronous*, facilitating communication among participants across time and space. In our analysis of this discussion group, we documented how participants built on each other's comments through back-and-forth agreement and dissension. These interactions showed how participants took into account other people's feedback as they constructed their own. In other words, the mentoring activity in this FIMFiction.net discussion group was not a one-to-one exchange, but rather a highly interactive series of exchanges among multiple participants. These public, persistent interactions facilitated both the *acceleration* of knowledge sharing and learning and the *accretion* of knowledge within the community. This multi-person, highly interactive structure represents a hallmark of distributed mentoring [11].

**Implications**

Though it bears resemblance to the knowledge exchange processes found in social Q&A sites, crowdsourced feedback, and online critiquing communities, distributed mentoring is distinct from these processes due to its explicit focus on mentoring relationships that provide sustained instrumental and psychosocial support to individuals seeking to develop their skills in a particular domain of interest [15,25]. It departs from formal mentoring models in its focus on informal affinity spaces [20,21], and it pushes the boundaries of traditional mentoring models by showing how networked technologies can shape mentoring relationships in new ways [16]. Scholars' previous demarcations between mentorship and related activities [8, 14, 27, 40] do not fully account for the growth evident within fanfiction writing communities. Therefore, the current work holds important theoretical implications in its expanded view of mentoring that accounts for the affordances of networked technologies.

The findings from the current study demonstrate that it is possible for networked publics to facilitate a type of mentoring that is distinct from face-to-face mentoring. Qualities such as asynchronous communication, the ability to search for online content easily, and the public, persistent nature of community interactions support the seven attributes of distributed mentoring. The absence of these qualities in non-networked environments limits the forms that mentoring can take offline. Without the ability to communicate with many people across time and space and aggregate their individual pieces of feedback, mentoring in offline contexts lends itself to more traditional, one-to-one relationships, which are easier to maintain in non-networked environments.

With respect to the practical utility of distributed mentoring, the seven attributes could be used to inform our understanding of the social dynamics present in other online affinity spaces, such as DeviantArt (an art community), Ravelry (a knit and crotchet community), and Wikipedia. Distributed mentoring could also illuminate the underpinnings of the negative dynamics that emerge on sites that attract trolls, such as Reddit and Voat. Insight into both the positive and negative dynamics of networked spaces could be especially useful to designers of new online affinity spaces seeking to create supportive environments that encourage community and personal growth.

In addition, the seven attributes of distributed mentoring could be used to guide the design of learning experiences that take advantage of networked technologies to offer learners ongoing feedback on their progress towards particular learning goals. Importantly, this feedback would not come solely from their teacher and a couple of classmates, as is now the case in most formal learning environments. Instead, learners might receive feedback from students in other schools and even countries, and perhaps experts in the specific field of study. In light of our focus on fanfiction writing, the most obvious application of distributed mentoring would be subjects with a heavy focus on writing, such as Language Arts. However, the focus need not remain so narrow, as learners rely on mentorship in many domains, from physics and chemistry to music and drama. Regardless of domain, learners would benefit by receiving richer, more complex feedback than they would otherwise receive from a single classroom experience. While the institutional and structural constraints of many learning environments can make networked communication daunting, the potential benefits to learners and instructors are profound.

**Limitations and Future Directions**
This current study examines just two publicly available data sources—reader reviews and a group discussion—of a complex network that constitutes distributed mentoring, a network which also includes private messaging, phone calls, and other exchanges between authors and readers. Our entire study focused on just three fanfiction communities, limiting our ability to generalize the findings. Future work could examine evidence for distributed mentoring in a broader range of fandoms, including those dedicated to different genres, for example, games and sports. Beyond fan communities, we anticipate that distributed mentoring exists in other knowledge-sharing communities, such as Wikipedia, multiplayer games, and DIY sites. To investigate this possibility, future work could apply our coding scheme to document the type and prevalence of feedback generated by participants in these different online communities.

Based on our research so far, we suspect that distributed mentoring plays a positive role in authors' development as writers. However, we cannot make any definitive claims with our existing evidence. Future work should explore changes in the writing quality of fanfiction authors over time and the degree to which distributed mentoring contributes to these changes. We hypothesize that authors who experience a greater amount of distributed mentoring will improve the quality of their writing to a greater degree than authors who experience less distributed mentoring. To study this question, future work could extract stories, reviews, and associated metadata from fanfiction sites and investigate the relationship between distributed mentoring and writing quality (*e.g.,* grammar, reading level). Such an investigation would yield valuable knowledge about the extent to which distributed mentoring supports learning.

**CONCLUSION**
As affinity spaces and active sites of participatory culture, online fanfiction communities represent rich contexts for studying informal mentoring processes. Prior work [11] described the characteristics of these processes and how they are supported by networked technologies. Drawing on Hutchins' theory of distributed cognition, Campbell et al. [11] called this network-enabled form of collective feedback *distributed mentoring*. Distributed mentoring comprises seven attributes, each tied to the affordances of networked publics: *aggregation, accretion, acceleration, abundance, availability, asynchronicity,* and *affect*. In the current study, we extended this work by documenting the processes of distributed mentoring across a larger dataset, as well as by delving into its operation on a moment-by-moment basis. Specifically, we presented a thematic analysis of 4,500 reader reviews and an in-depth case study of a single fanfiction discussion group to provide systematic documentation of the seven attributes of distributed mentoring and their interrelationships. These complementary studies show how the seven attributes work together to guide fanfiction authors as they seek to develop their writing skills. They also help us to further distinguish distributed mentoring from traditional models of mentoring as well as online forms of peer production like social Q&A sites and crowdsourced feedback. Distributed mentoring holds potential for supporting learning in a variety of formal and informal learning contexts.

**ACKNOWLEDGMENTS**

We thank the reviewers for their thoughtful and detailed feedback which greatly improved the paper. We also extend gratitude to our participants in this study.

### REFERENCES

1. Muhammad A. Ahmad, David Huffaker, Jing Wang, Jeff Treem, Dinesh Kumar, Marshall S. Poole, and Jaideep Srivastava. 2010. The many faces of mentoring in an MMORPG. In *Proc. of the IEEE Conference on Social Computing* (SocialCom, '10), 270-275.

2. Rebecca W. Black. 2006. Language, culture, and identity in online fanfiction. *E-Learning* 3, 2, 170-184.

3. Rebecca W. Black. 2007. Digital design: English language learners and reader reviews in online fiction. In *A new literacies sampler*, Michele Knobel and Colin Lankshear (eds.). Peter Lang, New York, NY, 115-136.

4. Rebecca W. Black. 2008. *Adolescents and Online Fan Fiction*. Peter Lang, New York, NY.

5. Tom Boellstorff, Bonnie Nardi, Celia Pearce, and T. L. Taylor. 2012. *Ethnography and Virtual Worlds: A Handbook of Method*. Princeton University Press, Princeton.

6. Richard E. Boyatzis. 1998. *Transforming Qualitative Information: Thematic Analysis and Code Development*. Sage.

7. Dana Boyd. 2007. The role of networked publics in teenage social life. In *Youth, identity, and digital media*. MIT Press, Cambridge, MA, 119-142.

8. Barry Bozeman, and Mary K. Feeney, M. 2007. Toward a useful theory of mentoring: A conceptual analysis and critique. *Admin & Soc,* 39, 6, 719-739.

9. Amy Bruckman. 1998. Community Support for Constructionist Learning. *CSCW* 7, 47-86.

10. Susan Bryant, Andrea Forte and Amy Bruckman. 2005. Becoming Wikipedian: Transformation of Participation in a Collaborative Online Encyclopedia. *Proceedings of ACM GROUP* (GROUP '05), 1-10.

11. Julie Campbell, Cecilia Aragon, Katie Davis, Sarah Evans, Abigail Evans, and David Randall. 2016. Thousands of positive reviews: distributed mentoring in online fan communities. In *Proc ACM Conference on Computer Supported Cooperative Work* (CSCW '16), 691-704. http://dx.doi.org/10.1145/2818048.2819934

12. Kelly Chandler-Olcott and Donna Mahar. 2003. Adolescents' *anime*-inspired "fanfictions": An exploration of multiliteracies. *J Adolesc Adult Lit.* 46, 7, 556-566.

13. Kwangsu Cho, Christian D. Schunn, and Davida Charney. 2006. Commenting on writing typology and perceived helpfulness of comments from novice peer reviewers and subject matter experts. *Written Comm*, 23, 3, 260-294.

14. Phillip Dawson. 2014. Beyond a definition: Toward a framework for designing and specifying mentoring models. *Ed Researcher,* 43, 3, 137-145.

15. Lillian T. Eby, Jean E. Rhodes, and Tammy D. Allen. 2008. Definition and evolution of mentoring. In *The Blackwell Handbook of Mentoring: A Multiple Perspectives Approach*, 7-20.

16. Ellen A. Ensher, Christian Heun, and Anita Blanchard. 2003. Online mentoring and computer-mediated communication: New directions in research. *J of Vocational Behavior,* 63, 2, 264-288.

17. Joseph L. Fleiss, Bruce Levin, and Myunchee Cho Paik. 2013. *Statistical methods for rates and proportions*. John Wiley & Sons.

18. Deborah A. Fields, Katarina Pantic, and Yasmin B. Kafai. 2015. "I have a tutorial for this": the language of online peer support in the scratch programming community. In *Proceedings of the ACM Conference on Interaction Design and Children* (IDC '15), 229-238.

19. FIMFiction.net. FIMFiction.net Statistics. Retrieved May 23, 2016 from http://www.fimfiction.net/statistics

20. James Paul Gee. 2004. *Situated Language and Learning: A Critique of Traditional Schooling*. Routledge, New York, NY.

21. James Paul Gee. 2007. *What Video Games Have to Teach Us About Learning and Literacy* (Rev. and updated ed.). Palgrave Macmillan, New York, NY.

22. James Paul Gee. 2013. *The Anti-Education Era: Creating Smarter Students Through Digital Learning*. Palgrave Macmillan, New York, NY.

23. F. Maxwell Harper, Daniel Moy, and Joseph A. Konstan. 2009. Facts or friends?: distinguishing informational and conversational questions in social Q&A sites. In *Proceedings of the SIGCHI Conference on Human Factors in Computing Systems* (CHI '09), 759-768.

24. Shirley Brice Heath, and Brian V. Street. 2008. *On Ethnography: Approaches to Language and Literacy Research. Language & Literacy (NCRLL)*. Teachers College Press, New York, NY.

25. Monica C. Higgins, and Kathy E. Kram. 2001. Reconceptualizing mentoring at work: A developmental network perspective. *Academy of Mgmt Review*, 26, 2, 264-288.
26. Julie Hui, Amos Glenn, Rachel Jue, Elizabeth Gerber, and Steven Dow. 2015. Using anonymity and communal efforts to improve quality of crowdsourced feedback. In *Third AAAI Conf on Human Computation and Crowdsourcing* (HCOMP '15), 72-82.
27. Russel L. Huizing. 2012. Mentoring together: A literature review of group mentoring. *Mentoring & Tutoring*, 20,1, 27-55.
28. Beth Humberd and Elizabeth Rouse. 2015. Seeing you in me and me in you: Personal identification in the phases of mentoring relationships. *Academy of Mgmt Review*, 41, 3, 435-455.
29. David Marshall Hunt, and Carol Michael. 1983. Mentorship: A career training and development tool. *Academy of Mgmt Review,* 8, 3, 475-485.
30. Edwin Hutchins. 1995. *Cognition in the Wild*. MIT Press, Cambridge, MA.
31. Edwin Hutchins and Tove Klausen. 1996. Distributed cognition in an airline cockpit. In *Cognition and communication at work*, Yrjö Engeström and David Middleton (eds.). Cambridge University Press, 15-34.
32. Beverly J. Irby, and Jennifer Boswell. 2016. Historical print context of the term, "mentoring." *Mentoring & Tutoring,* 24, 1, 1-7.
33. Mizuko Itō. 2010. *Hanging Out, Messing Around, and Geeking Out: Kids Living and Learning with New Media*. MIT Press, Cambridge, MA.
34. Mizuko Ito, Kris Gutiérrez, Sonia Livingstone, Bill Penuel, Jean Rhodes, Katie Salen, Juliet Schor, Julian Sefton-Green and S. Craig Watkins. 2013. *Connected learning: An agenda for research and design*. Digital Media and Learning Research Hub.
35. Henry Jenkins. 1992. *Textual Poachers: Television Fans & Participatory Culture.* Routledge, New York, NY.
36. Henry Jenkins. 2006. *Convergence Culture: Where Old and New Media Collide*. New York University Press.
37. Henry Jenkins. 2009. *Confronting the Challenges of Participatory Culture: Media Education for the 21st Century*. MIT Press, Cambridge, MA.
38. Debi Khasnabis, Catherine H. Reischl, Melissa Stull, Timothy Boerst. 2013. Distributed mentoring: Designing contexts for collective support of teacher learning. *English Journal*, 102, 3, 71.
39. Kathy E. Kram. 1985. *Mentoring at work: Developmental relationships in organizational life* (Organizational behavior and psychology series). Glenview, Ill.: Scott, Foresman.
40. Jonathan Kroll. 2016. What is meant by the term group mentoring? *Mentoring & Tutoring,* 24, 1, 44-58.
41. Jean Lave and Etienne Wenger. *Situated learning: Legitimate peripheral participation*. Cambridge University Press, New York, 1991.
42. Daniel J. Levinson, Charlotte N. Darrow, Maria H. Levinson, Edward B. Klein, and Braxton McKee. 1978. *Seasons of a man's life*. New York: Academic Press.
43. Jennifer Marlow and Laura Dabbish. 2014. From rookie to all-star: professional development in a graphic design community of practice. In *Proc ACM Conference on Computer Supported Cooperative Work & Social Computing* (CSCW '14), 922-933.
44. Sharan B. Merriam. 2009. *Qualitative research: A guide to design and implementation*. Jossey-Bass, San Francisco, CA.
45. Matthew B. Miles and A. Michael Huberman. 1994. *Qualitative Data Analysis: An Expanded Sourcebook.* Sage Publications, Thousand Oaks, CA.
46. Michael J. Moore, Tadashi Nakano, Akihiro Enomoto, and Tatsuya Suda. 2012. Anonymity and roles associated with aggressive posts in an online forum. *Comp in Hum Behav*, 28, 3, 861-867.
47. Meredith Ringel Morris, Kori Inkpen Quinn, and Gina Venolia. 2014. Remote shopping advice: enhancing in-store shopping with social technologies. In *Proc ACM Conference on Computer Supported Cooperative Work & Social Computing* (CSCW '14), 662-673.
48. Stephen C. Scogin and Carol L. Stuessy. 2015. Encouraging greater student inquiry engagement in science through motivational support by online scientist-mentors. *Sci Edu*, 99, 2, 312-349.
49. Peter Smagorinsky. 2008. The method section as conceptual epicenter in constructing social science research reports. *Written Comm*, 25, 3, 389-411.
50. Miranda M. W. C. Snoeren, Ragna Raaijmakers, Theo J. H. Niessen, and Tineke A. Abma. 2016. Mentoring with(in) care: A co-constructed auto-ethnography of mutual learning. *J of Org Behavior,* 37, 1, 3-22.
51. Heidrun Stoeger, Xiaoju Duan, Sigrun Schirner, Teresa Greindl, and Albert Ziegler. 2013. The effectiveness of a one-year online mentoring program for girls in STEM. *Comp & Edu*, 69, 408-418.
52. Anbang Xu and Brian Bailey. 2012. What do you think?: a case study of benefit, expectation, and interaction in a large online critique community. In *Proc ACM Conference on Computer Supported Cooperative Work* (CSCW '12), 295-304.
53. José Zagal and Amy Bruckman. 2010. Designing Online Environments for Expert/Novice Collaboration: Wikis to Support Legitimate Peripheral Participation. *Convergence* 16, 4, 451-470.